\documentclass[twocolumn,superscriptaddress,showpacs,preprintnumbers,amsmath,amssymb]{revtex4}

\usepackage{graphicx}
\usepackage{dcolumn}
\usepackage{bm}
\usepackage{color}
\usepackage{soul}

\begin{document}

\title{Signature of a highly spin polarized resonance state at Co$_2$MnSi(001)/Ag(001) interfaces.}

\author{Christian Lidig}
	\affiliation{Institut f\"ur Physik, Johannes Gutenberg-Universit\"at Mainz, 55099 Mainz, Germany}    
\author{Jan Min\'ar}
	\affiliation{New Technologies-Research Center, University of West Bohemia, Univerzitni 8, 306 14 Pilsen, Czech Republic}
\author{J\"urgen Braun}
	\affiliation{Department Chemie, Ludwig-Maximilians-Universit\"at M\"unchen, Butenandtstrasse 11, 81377 M\"unchen, Germany}
\author{Hubert Ebert}
	\affiliation{Department Chemie, Ludwig-Maximilians-Universit\"at M\"unchen, Butenandtstrasse 11, 81377 M\"unchen, Germany}
\author{Andrei Gloskovskii}
	\affiliation{Deutsches Elektronen-Synchrotron DESY, 22603 Hamburg, Germany}
\author{Alexander Kronenberg}
	\affiliation{Institut f\"ur Physik, Johannes Gutenberg-Universit\"at Mainz, 55099 Mainz, Germany}     
\author{Mathias Kl\"aui}
	\affiliation{Institut f\"ur Physik, Johannes Gutenberg-Universit\"at Mainz, 55099 Mainz, Germany}
\author{Martin Jourdan}
	\affiliation{Institut f\"ur Physik, Johannes Gutenberg-Universit\"at Mainz, 55099 Mainz, Germany}   

\date{\today}

\begin{abstract}
We investigated interfaces of halfmetallic Co$_2$MnSi(100) Heusler thin films with Ag(100), Cr(100), Cu and Al layers relevant for spin valves by high energy x-ray photoemission spectroscopy (HAXPES). Experiments on Co$_2$MnSi samples with an Ag(100) interface showed a characteristic spectral shoulder feature close to the Fermi edge, which is strongly diminished or suppressed at Co$_2$MnSi (100) interfaces with the other metallic layers. This feature is found to be directly related to the Co$_2$MnSi(100) layer, as reflected by control experiments with reference non-magnetic films, i.\,e.\,without Heusler layer. By comparison with HAXPES calculations, the shoulder feature is identified as originating from an interface state related to a highly spin polarized surface resonance of Co$_2$MnSi (100).
\end{abstract}

\maketitle

\section{Introduction}
The magnitude of the spin polarization of ferromagnetic materials is a key property for their application in spin transport-based electronics\cite{Wolf01}. However, it is not the bulk, but the surface or interface electronic properties, which are relevant for most applications. Investigating epitaxial thin films of the Heusler compound Co$_2$MnSi \cite{Gal02,And16,Gra11,Ish95} by spin resolved UV-photoemission spectroscopy \cite{Sch15,Hah11,Kol12} and spin-integrated high energy x-ray photoemission spectroscopy (HAXPES), recently a high spin polarization was observed at room temperature in a wide energy range below the Fermi energy \cite{Jou14}. This resonance is related to a surface resonance in the majority electron spin band extending deep into the bulk of the material \cite{Jou14, Bra15}. 
Correspondingly, Co$_2$Fe$_x$Mn$_{1-x}$Si / Ag / Co$_2$Fe$_x$Mn$_{1-x}$Si spin valves show giant magnetoresistance (GMR) values up to 170~\% at low temperatures \cite{Sak12}. However, the use of alternative non-magnetic spacer layers like thin Cr always resulted in strongly reduced GMR values  \cite{Yak06} with unclear origin. In order to understand this behavior, we investigated Co$_2$MnSi interfaces with several metals typically used as spacer layers for spin valves. The high photon energy of HAXPES (typically $6-8$~keV) corresponds to a relatively large inelastic mean free path of the photoelectrons of $\approx 10$~nm \cite{Tan11}, which allows to probe electronic bulk states and buried interfaces. However, the small scattering cross sections in HAXPES result in general in a relatively small photoemission intensity. As spin-filtering typically is associated with an additional reduction of the count rates by orders of magnitude, it is not available as a standard method \cite{Koz16}. 

Our previous HAXPES investigations (6~keV photon energy) of Co$_2$MnSi(100)/AlO$_x$~(2~nm) thin films showed a characteristic shoulder feature near the Fermi edge. This feature is, based on comparison with HAXPES calculations considering a bare Co$_2$MnSi(100) surface, associated with the highly spin-polarized surface resonance \cite{Jou14, Bra15} (Fig.\,1).
\begin{figure}[htb]
\centering
		\includegraphics[width=0.90\columnwidth]{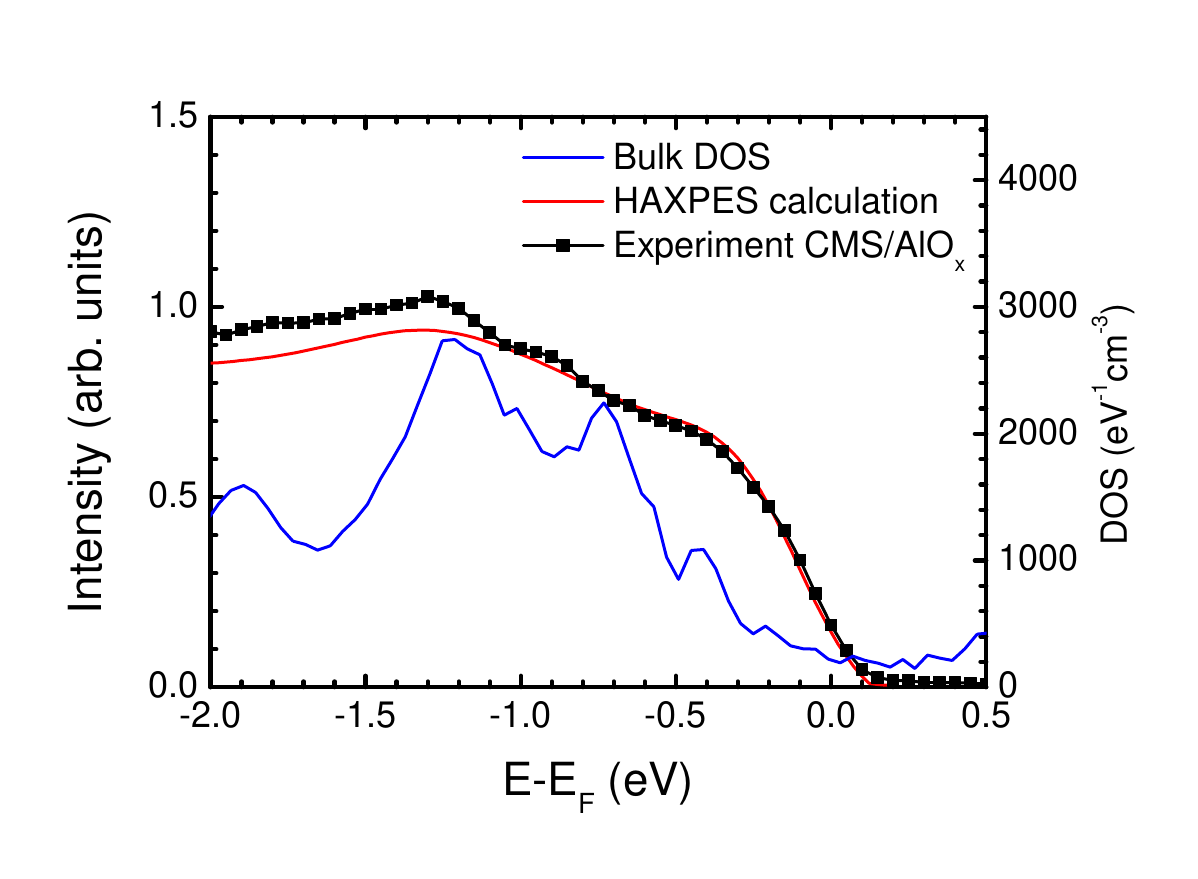}
	\caption{HAXPES intensity obtained probing a Co$_2$MnSi(100) / AlO$_x$ (2~nm) sample with 6~keV excitation energy in comparison with a HAXPES calculation assuming a bare Co$_2$MnSi(100) surface. Additionally, the calculated bulk density of states of Co$_2$MnSi is shown \cite{Jou14}.}
\end{figure}
Here we propose to use this HAXPES shoulder feature as a signature of high spin polarisation also for buried metal interfaces. First, the previously implied association of the AlO$_x$~(2~nm) capping layer with a vacuum interface is reconsidered by investigations of Co$_2$MnSi(100) / Al (2~nm) / AlO$_x$ (2~nm) trilayers, before other Co$_2$MnSi(100) / metal (2~nm) / AlO$_x$ (2~nm) trilayers are discussed.

\section{Sample preparation and characterization}
The multilayer samples were prepared by radio-frequency (RF)-sputtering at room-temperature. Epitaxial Co$_2$MnSi(100) layers with a typical thickness of 20\,nm were deposited directly on MgO(100) substrates. After an in-situ annealing process at 550$^\circ$C, L2$_1$ order was obtained and verified by X-ray diffraction and in-situ electron diffraction (RHEED). The arrangement of the RHEED spots on semi-circles as shown in the upper panel of Fig.\,2 indicates scattering on a 2-dimensional surface, which is consistent with the generally very smooth morphology of this type of Heusler thin films \cite{Her09}. The alternating intensity of neighboring RHEED spots is characteristic for a L2$_1$ ordered surface.  Co-Mn swapping disorder (B2 structure) would result in a vanishing of the weaker spots. After allowing the sample to cool down within 45~min to $\simeq 100^\circ$C in ultra high vacuum ($\simeq 10^{-10}$~mbar), either Al, Cr, Cu, or Ag layers (2~nm) were deposited in-situ on top of the Co$_2$MnSi thin films. The RHEED investigations of these layers demonstrated epitaxial growth of Ag(100) (Fig.\,2, lower panel) and Cr(100), whereas Al and Cu layers showed diffuse RHEED images indicating polycrystalline growth. 
Finally, all samples were capped with 2\,nm AlO$_x$ as protection against oxidation.
\begin{figure}[htb]
\centering
		\includegraphics[width=0.60\columnwidth]{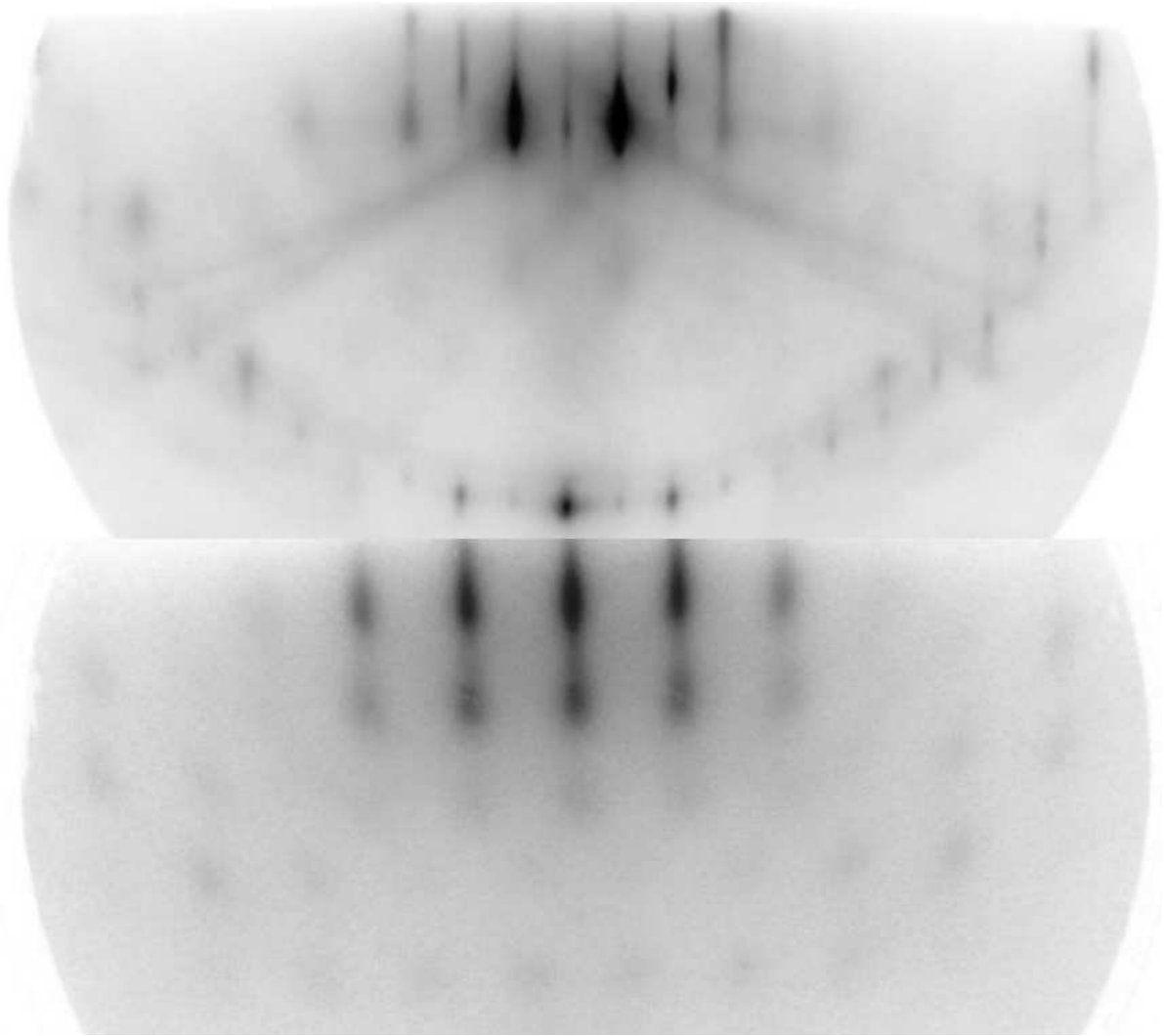}
	\caption{The RHEED image of a Co$_2$MnSi(100) thin film (upper panel) demonstrates the surface crystallographic L2$_1$ order of the Heusler thin film, on which an Ag (100) layer was grown epitaxially as demonstrated by the second RHEED image (lower panel).}
\end{figure}

\section{Experimental HAXPES investigations}
The multilayer samples described above were investigated ex-situ by HAXPES with a photon energy of 6~keV at beamline P09 of the Deutsches Elektronen-Synchrotron (DESY), Hamburg \cite{Glo12}. The SPECS Phoibos 225HV spectrometer was operated in the transmission mode of the transfer lens which is optimized for high transmission at the expense of angular information. The entrance slit of the spectrometer with a size of 3x30 mm was aligned parallel to the in-plane [110]-direction of the Co$_2$MnSi epitaxial thin film. The angular acceptance was $\pm 15^{\rm o}$ and $\pm 1.55^{\rm o}$ parallel and perpendicular to the entrance slit, respectively. The angle of x-ray incidence was $5^{\rm o}$ with respect to the surface normal. All experiments discussed were performed at room temperature.

In principle, the shoulder feature in the HAXPES intensity near the Fermi edge shown in Fig.\,1 could result from metallic states of an underoxidized AlO$_x$ capping layer. For the investigation of such effects we compare the HAXPES spectrum shown in Fig.\,3 obtained when measuring a Co$_2$MnSi(100) / Al~(2~nm) / AlO$_x$~(2~nm) trilayer with the spectrum from the Co$_2$MnSi(100) / AlO$_x$~(2~nm) sample shown in Fig.\,1. It is obvious that with the insertion of the additional metallic Al layer the shoulder feature is completely suppressed, ruling out a simple capping layer related origin and thus corroborating the highly spin polarized surface resonance based theoretical interpretation given in our previous work \cite{Jou14,Bra15}. 

Next, the influence of various potential spacer layers of GMR devices on the HAXPES shoulder feature is studied by comparing all results shown in Fig.\,3. Investigating the fully epitaxial Co$_2$MnSi(100) / Ag(100) interface the obtained HAXPES intensities are very similar to the Co$_2$MnSi(100) / AlO$_x$ case, which suggests the conversion of the surface resonance into an interface state for these samples. 

In the HAXPES data obtained investigating the Co$_2$MnSi(100) / Cu (polycrystalline) interface the shoulder feature is strongly diminished, but still observable. Investigating epitaxial Co$_2$MnSi(100) / Cr(100) interfaces, as in the case of the Co$_2$MnSi(100) / Al (polycrystalline) interfaces described already above, no indication at all of the shoulder feature related to the surface resonance of Co$_2$MnSi(100) is found. 

\begin{figure}[htb]
\centering
		\includegraphics[width=1.00\columnwidth]{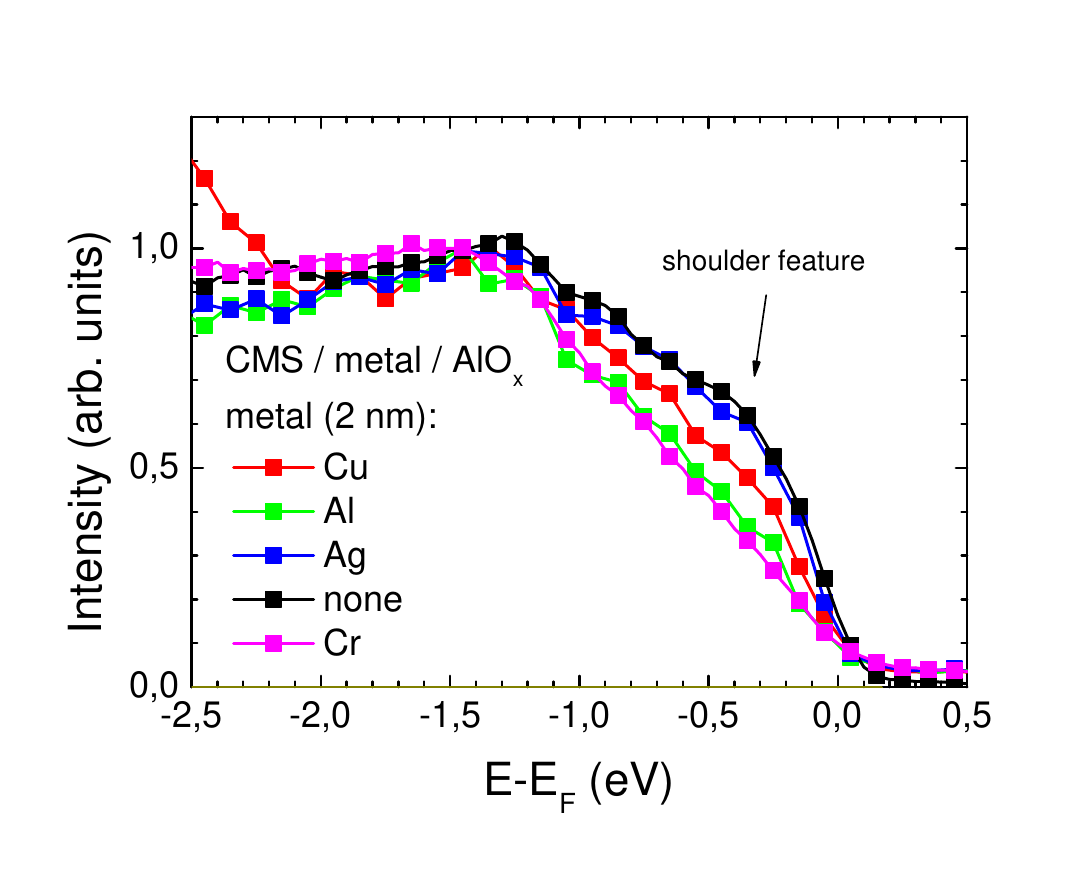}
	\caption{HAXPES intensities obtained with a photon energy of 6~keV (energy resolution $\simeq 300$~meV) investigating different Co$_2$MnSi(001) / metal (2~nm) / AlO$_x$ (2nm) samples. The arrow marks the position of the discussed highly spin polarized surface resonance.}
\end{figure}

\begin{figure}[htb]
\centering
	\includegraphics[width=1\columnwidth]{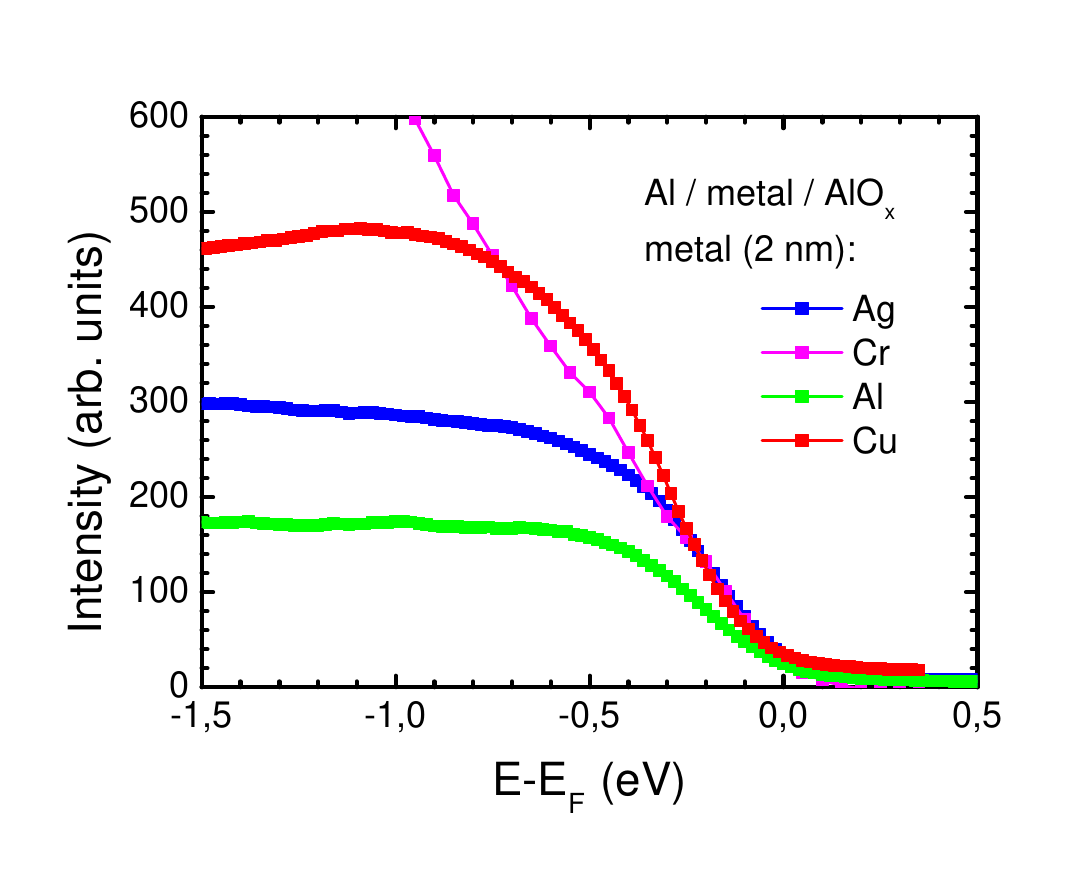}
		\caption{HAXPES spectra (photon energy 6~keV, energy resolution 300~meV) obtained investigating MgO / Al(10~nm) / metal (2nm)/ AlO$_x$ multilayers with different metal layers as indicated in the figure.}
\end{figure}

For the interpretation of the HAXPES results described above, it is important to estimate the contribution of the bulk electronic states of the 2~nm metal layers to the spectra. Experimentally, this is possible by investigating the metallic top layers seperately without a Heusler thin film below. As HAXPES of 2~nm metal plus 2~nm AlO$_x$ capping deposited directly on MgO substrates resulted in charging of the samples, we investigated MgO / Al(10~nm) / metal (2~nm) / AlO$_x$ (2~nm) samples. In Fig.\,4 spectra obtained with identical experimental settings are shown allowing for a comparison of the intensities measured at the Fermi edge. In the cases of 2~nm Ag and Cu metal layers as well as with metallic Al below the AlO$_x$ capping layer similar intensities were obtained. In the case of the 2~nm Cr layer the intensity $\simeq 1.5$~eV below the Fermi-energy was much higher, which is to be expected due to the d-electron character at the Fermi level of this compound. Relating these intensities to the appearance or distinctiveness of the shoulder feature in HAXPES of the Heusler / metal (2~nm) / AlO$_x$(2~nm) samples, it is obvious that the additional intensity near the Fermi level in the case of Co$_2$MnSi(001) / Ag(001) is not caused by bulk contributions of the Ag layer itself, rendering from a purely experimental point of view the observation of a specific surface resonance derived interface state possible.\\ 

\section {Band structure and HAXPES calculations} 
To gain deeper insight into the electronic origin of the shoulder feature discussed above and its relation to the highly spin polarized surface resonance of bare Co$_2$MnSi(001), band structure and HAXPES calculations for the epitaxial Co$_2$MnSi(001)/Ag(001) system were performed. The experimentally investigated Co$_2$MnSi(001) / Al (or Cu) interfaces could not be considered theoretically, because the polycrystalline structure of the normal metals involved is not describable by our theoretical models. Also the HAXPES intensity for epitaxial Co$_2$MnSi(001) / Cr(001) interfaces was not calculated, in this case due to the unknown magnetic order of the 2~nm Cr layer, which is a complex topic by itself \cite{Zab99}.   

The self-consistent electronic structure calculations were performed within the framework of ab-initio spin-density functional theory by use of the local density approximation. The electronic structure of 6 monolayers Ag on Co$_2$MnSi(001) was computed for a semi-infinite system using the fully relativistic screed Korringa-Kohn-Rostoker method as implemented in the Munich SPR-KKR package \cite{SPR-KKR7.7,EKM11}. All technical details can be found in \cite{Jou14,Bra15}. Corresponding to the experiments all calculations were performed assuming
in-plane magnetized samples. Self-consistent potentials were used as input for angle-integrated HAXPES calculations of the Co$_2$MnSi(001)/Ag(001) surface. We accounted for the surface by use of a Rundgren-Malmstr\"om barrier potential \cite{MR80}, which was included as an additional layer in the photoemission analysis. This procedure is described in detail in Ref.\,\cite{NBF+11} and accounts quantitatively for the energetics and dispersion of all surface-related features. Furthermore, the relative intensities of surface states and resonances are accounted
for by calculating the corresponding matrix elements in the surface regime. As the LSDA does not consider the finite life-time of the initial state, its effect was treated phenomenologically by including an imaginary part of the potential (0.03~eV). The impurity scattering of the final state and its inelastic mean free path was modeled by the imaginary part of the inner potential (10.0~eV) \cite{Bra96}. The chosen value results in a photocurrent that decays exponentially by one order of magnitude within the first 35 atomic layers.
This way the bulk sensitivity is guaranteed.

In Fig.\,5 we show the calculated spin-resolved total density of states (DOS) for 6 monolayers of Ag on top of the semi-infinite Co$_2$MnSi(001) surface, which was obtained by adding the averaged DOS of the Ag layers to the bulk DOS of Co$_2$MnSi(001). At the interface with Ag the spin polarisation at the Fermi level is slightly reduced due to interface and surface states found in the minority channel. However, as shown in Fig.\,5, the Ag states are located far below the Fermi level and therefore contribute to the spectral features at the Fermi level not significantly.
\begin{figure}[htb]
\centering
\includegraphics[width=0.8\columnwidth]{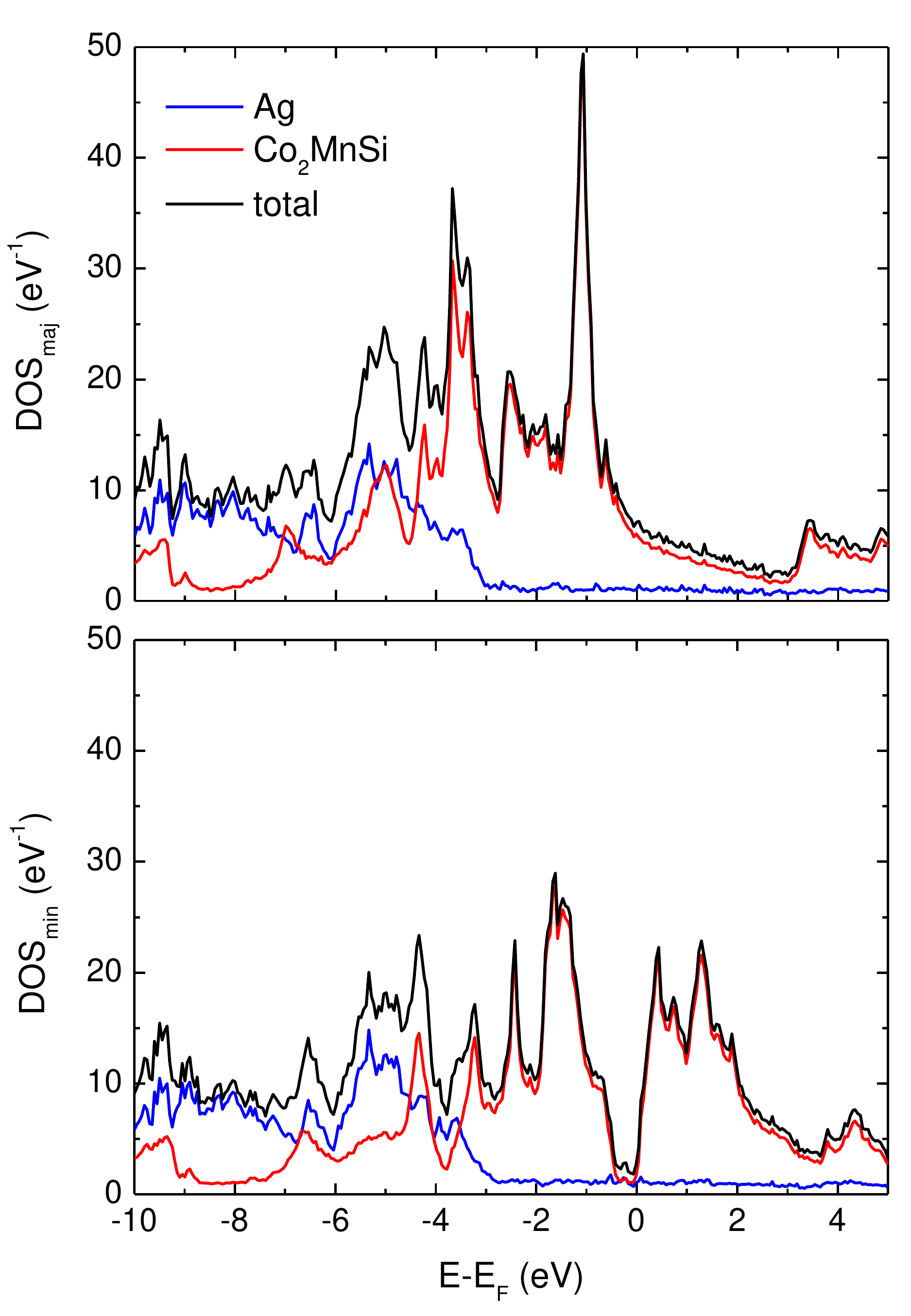}
	\caption{Spin resolved density of states (DOS) for the Co$_2$MnSi(001)/Ag(001) system. Upper panel: DOS of the majority charge carriers. Lower panel: DOS of the minority charge carriers. The total density of states obtained by averaging the contributions of Co$_2$MnSi(001) and Ag(001) is shown in black. Additionally, the Co$_2$MnSi-bulk-DOS (red) and the Ag-DOS (green) obtained by averaging 6 epitaxial Ag monalayers on top of the Heusler thin film are shown.}
\end{figure}  

As shown in our previous publications \cite{Jou14,Bra15}, the close to 100\% spin polarisation obtained by UV-photoemission experiments is ascribed to a surface resonance in the majority channel. As such resonances are typically hidden in the bulk continuum, they cannot be assigned directly from ground state-electronic calculations. However, matrix element effects enhance the spectral intensity of these features significantly.
Similar states can be found by analyzing the ground state Bloch spectral function. These states are localized at the interface between Co$_2$MnSi(001) and Ag and are situated slightly above the Fermi level with high Co partial character. However, these interface resonances show very low spectral weight and it is difficult to identify them because they are found not only in the band gap but also disperse in the bulk continuum. Thus, additionally to our ground state electronic structure calculations, we performed angle-integrated HAXPES one-step model calculations, which are shown in Fig.\,6 in comparison with the corresponding experimental data. The good agreement of both data sets, specifically concerning the spectral shoulder at $E-E_F\simeq -0.5$~eV, is obvious. From this we conclude that a highly spin-polarized surface resonance exists at the Co$_2$MnSi(001)/Ag(001) interface as well. 
 
\begin{figure}[htb]
\centering
		\includegraphics[width=1.00\columnwidth]{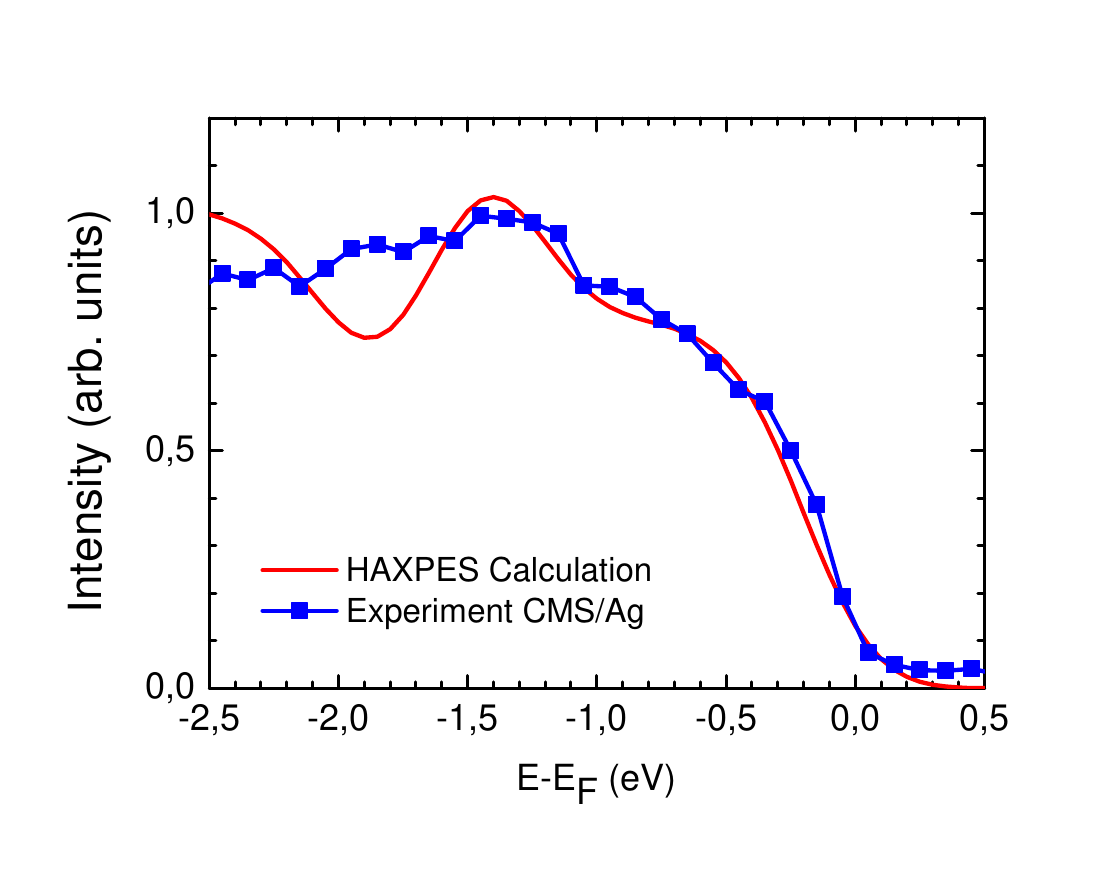}
	\caption{Comparison of the experimentally and theoretically obtained HAXPES intensities (photon energy 6~keV, energy resolution $\simeq 300$~meV ) investigating a Co$_2$MnSi(001) / Ag(001) (2~nm) / AlO$_x$ (2nm) sample.}
\end{figure}  


The resonance discussed above is a specific feature of the fully epitaxial Co$_2$MnSi(001)/Ag(001) system. A necessary precondition for the development of interface resonances as discussed above is their coupling to dispersive states, which disappear in the polycrystalline case. This is consistent with the suppression of the HAXPES shoulder feature observed if the epitaxial Ag(001) layer is replaces by polycrystalline Al or Cu. In general, also the energetic position of the surface resonance depends on the specific metal deposited on top of the Co$_2$MnSi layer. This is because the electronic structure of the interface region depends on the adsorbate, and the resonance by definition has a significant bulk contribution. Thus the surface resonance is sensitive in its dispersional behavior to the corresponding bulk states.
Experimentally, we also observed for the epitaxial Co$_2$MnSi(001)/Cr(001) system a full suppression of the related HAXPES shoulder feature. However, as already mentioned above, the magnetic structure of the 2~nm thin Cr(001) on Co$_2$MnSi(001) is unknown. Thus no attempt to calculate the corresponding HAXPES spectrum was made.

\section{Summary}
By experimental HAXPES investigations of Co$_2$MnSi(100) / Ag(100)(2~nm) / AlO$_x$(2~nm) trilayer samples and corresponding calculations, evidence for a highly spin polarized interface state close to the Fermi level of the Heusler compound is found. The corresponding shoulder like HAXPES feature is diminished for interfaces with polycrystalline Cu and vanishes completely for interfaces with polycrystalline Al as well as with epitaxial Cr. These observations are consistent with the possibility to realize large GMR effects with Heusler / Ag interfaces, but not with alternative non-magnetic spacer layers and suggest a relation of the GMR with the existence/ non-existence of a spin polarized interface state.

\textit{ Financial  support  by  the  the  German  Research   Foundation (DFG) via project Jo404/9-1 is acknowledged. Funding for   the HAXPES instrument at beamline P09 by the Federal Ministry of   Education and Research (BMBF) under contracts 05KS7UM1 and 05K10UMA   with Universit\"at Mainz; 05KS7WW3, 05K10WW1 and 05K13WW1 with   Universit\"at W\"urzburg is gratefully acknowledged. We further thank for the support from CEDAMNF (CZ.02.1.01/0.0/0.0/15\_003/0000358) of Czech ministry MSMT. }

\end{document}